\documentclass{PoS}
\newcommand{\be}{\begin{eqnarray}}
\newcommand{\ee}{\end{eqnarray}}
\usepackage{epsfig}
\newcommand{\Ree}{{\rm Re}}
\newcommand{\Imm}{{\rm Im}}
\newcommand\nn{\nonumber}
\newcommand{\Tr}{{\rm Tr}}

\title{How the Quark Number fluctuates in QCD at small chemical potential}

\ShortTitle{Quark Number fluctuations in QCD}

\author{M.P. Lombardo\\
         INFN-Laboratori Nazionali de Frascati\\
         I-00044, Frascati (RM)\\
         Italy\\
         E-mail: \email{mariapaola.lombardo@lnf.infn.it}}
\author{\speaker{K. Splittorff}\\
         The Niels Bohr Institute \\
         Blegdamsvej 17, DK-2100 Copenhagen \\ Denmark\\
         E-mail: \email{split@nbi.dk}}
\author{J.J.M. Verbaarschot\\
         State University of New York\\
         Department of Physics and Astronomy\\
         Stony Brook, NY 11794-3800, USA\\
         E-mail: \email{verbaarschot@cs.physics.sunysb.edu}}

\abstract{We discuss the distribution of the quark number over the
  gauge fields for QCD at nonzero quark chemical potential. As the
  quark number operator is non-hermitian, the distribution is over the
  complex plane. Moreover, because of the fermion determinant, the
  distribution is not real and positive. The computation is carried
  out within leading order chiral perturbation theory and gives direct
  insight into the delicate cancellations that take place in
  contributions to the total baryon number.} 

\FullConference{The XXVIII International Symposium on Lattice Field Theory, Lattice2010\\
                June 14-19, 2010\\
                Villasimius, Italy}

\begin{document}

\section{Introduction}

First principle predictions for the QCD phase diagram would be of
great value as a benchmark for the international experimental 
heavy ion program. In principle, we know precisely how to proceed:
start from the grand canonical partition function, 
\be
Z_{1+1}
= \int dA \ {\det}^2(D+\mu\gamma_0+m) \ e^{-S_{\rm YM}},  
\ee
and study the baryon number as a function of the chemical potential, 
$\mu$, and temperature for quark masses $m$ (for notational
simplicity we consider the two flavor theory in these proceedings). As the
phase transitions occur in the non perturbative domain it is natural
to turn to lattice QCD. While we know exactly how to include the
chemical potential on the lattice \cite{HK,Kogut}, in practice, our
studies are limited by the fact that the fermion determinant at
nonzero values of the chemical potential becomes complex
\be
{\det}^2(D+\mu\gamma_0+m) =  |{\det}(D+\mu\gamma_0+m)|^2 e^{2i\theta}.
\ee
This {\sl sign problem} invalidates the standard Monte Carlo
method which is at the hart of lattice QCD. The sign problem is not
only hard because the average of the phase factor is exponentially
small \cite{Phase}, it is also challenging because much of our intuition for
statistical systems leads to wrong conclusions at non zero $\mu$. 
Most prominently, a probabilistic argument  leads one 
to conclude that the chiral condensate is continuous as a function of
the quark mass at $m = 0$ for any non zero value of $\mu$
\cite{Barbour}.   
Rather, as the solution \cite{OSV} of this Silver Blaze problem shows,  
it is imperative that we consider distribution functions that take 
complex values. Only through extreme complex oscillations of
the eigenvalue density of the Dirac operator is it possible to
understand how the discontinuity of the chiral 
condensate remains non zero in the presence of the chemical
potential. 
Here we will show \cite{LSV2} that the extreme complex oscillations take
place also in the chiral condensate and the baryon number 
and that they are essential to get the correct physical results. 

For simplicity we will here focus on the baryon number, which in a
single gauge field configuration is given by 
\be
\label{1}
n & \equiv &  
 \frac{d}{d\mu} \log\det(D+\mu\gamma_0+m). 
\ee
Our goal is to determine the distribution $\langle\delta(n-n')\rangle$
of the baryon number over the gauge fields. We will show that this
distribution takes complex values and that the extreme oscillations
are essential to obtain the correct average baryon number
\be
\langle n \rangle = \int dn' \ n' \ \langle\delta(n-n')\rangle. 
\ee
This direct insight into the sign problem also allow us to address how
complex Langevin works in this case.

In order to understand better the distribution
$\langle\delta(n-n')\rangle$ of the baryon number operator over the gauge
fields, let us first understand its first two moments. 
For the first moment there is no source of confusion: The average
quark number is the first moment
\be
\frac{1}{2}\frac{1}{Z_{1+1}} \ \frac{d}{d\mu}  \ Z_{1+1} = \langle n \rangle.
\ee
However, the second derivative with respect to $\mu$
\be
\label{nqSQ}
\frac{1}{4}\frac{1}{Z_{1+1}} \ \frac{d^2}{d\mu^2}  \ Z_{1+1} 
= \langle n^2 \rangle + \frac{1}{2} \langle\left(\frac{d n}{d\mu}\right) \rangle, 
\ee
is {\sl not} the second moment of the distribution
$\langle\delta(n-n')\rangle$. Rather, the average of the square of $n$
can be written as  
\be
\langle n^2 \rangle=\left. \frac{1}{Z} \ \frac{d}{d\mu_u}\frac{d}{d\mu_d} \  Z_{1+1} \right|_{\mu_u=\mu_d=\mu},
\ee
where we distinguished the chemical potentials for the two flavors and
differentiated with respect to each one before setting them equal.  

When we express the traces in terms of the eigenvalues, $z_{\,k}$, of
$\gamma_0(D+m)$  
\be
\frac{1}{2}\frac{1}{Z_{1+1}} \ \frac{d}{d\mu}  \ Z_{1+1} & = & \left\langle\sum_k \frac 1{z_k+\mu} \right\rangle \\
\frac{1}{4}\frac{1}{Z_{1+1}} \ \frac{d^2}{d\mu^2}  \ Z_{1+1}  & = & \left\langle\sum_{k , l} \frac 1{z_k+\mu} \frac
1{z_l+\mu} - \frac{1}{2}
\sum_k \frac 1{(z_k+\mu)^2}\right\rangle \nn\\
\left\langle n^2 \right\rangle &  = & \left\langle \sum_{k , l} \frac 1{z_k+\mu}  \frac
1{z_l+\mu}\right\rangle  = \left\langle\Big[\sum_k \frac 1{z_k+\mu} \Big]^2 \right\rangle,  \nn
\ee
it becomes obvious that (\ref{nqSQ}) is {\sl not} the
average of a square and in particular it is {\sl not} the second moment of 
the distribution of $n$ over the gauge fields. 
The distribution $\langle\delta(n-n')\rangle$ is nevertheless of great
interest since it gives direct insights in the sign problem.

As a final point before we turn to the results, note that the
quark number takes complex values  
\begin{equation}
n(\mu)^*=\left(\Tr\frac{\gamma_0}{D+\mu\gamma_0+m}\right)^*
=-n(-\mu). 
\end{equation}
Hence, the distribution $\langle\delta(n-n')\rangle$ is in the 
complex $n$ plane  
\be
P_{n}(x,y)\equiv\left\langle
\delta\left(x-{\rm Re}[n]\right)
\delta\left(y-{\rm Im}[n]\right)\right\rangle,
\ee
and the average baryon number is given by the integral of
$(x+iy)$ weighted by the distribution $P_{n}(x,y)dxdy$.   

\begin{figure}[t!]
\unitlength1.0cm
\epsfig{file=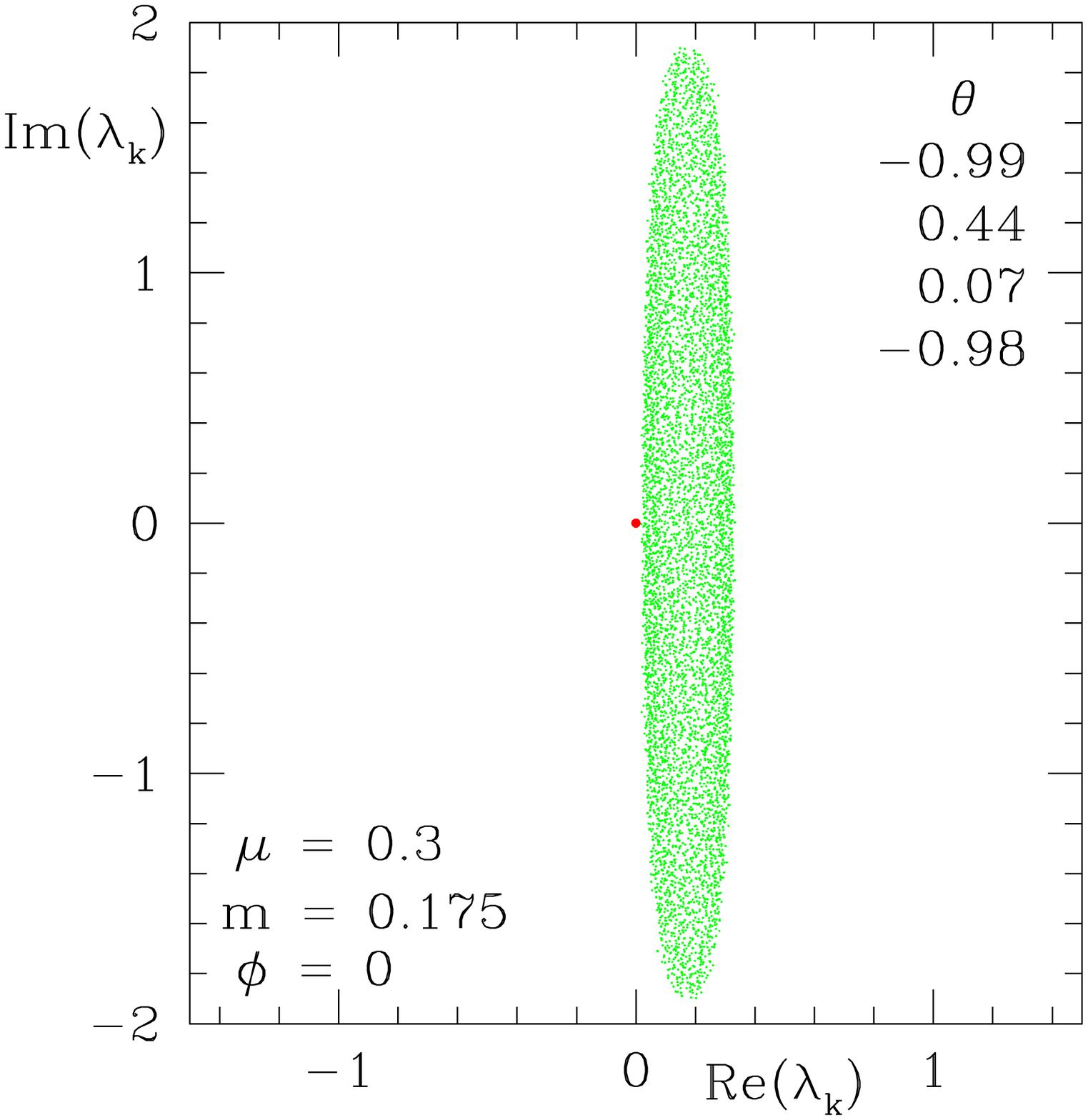,clip=,width=7cm}
\epsfig{file=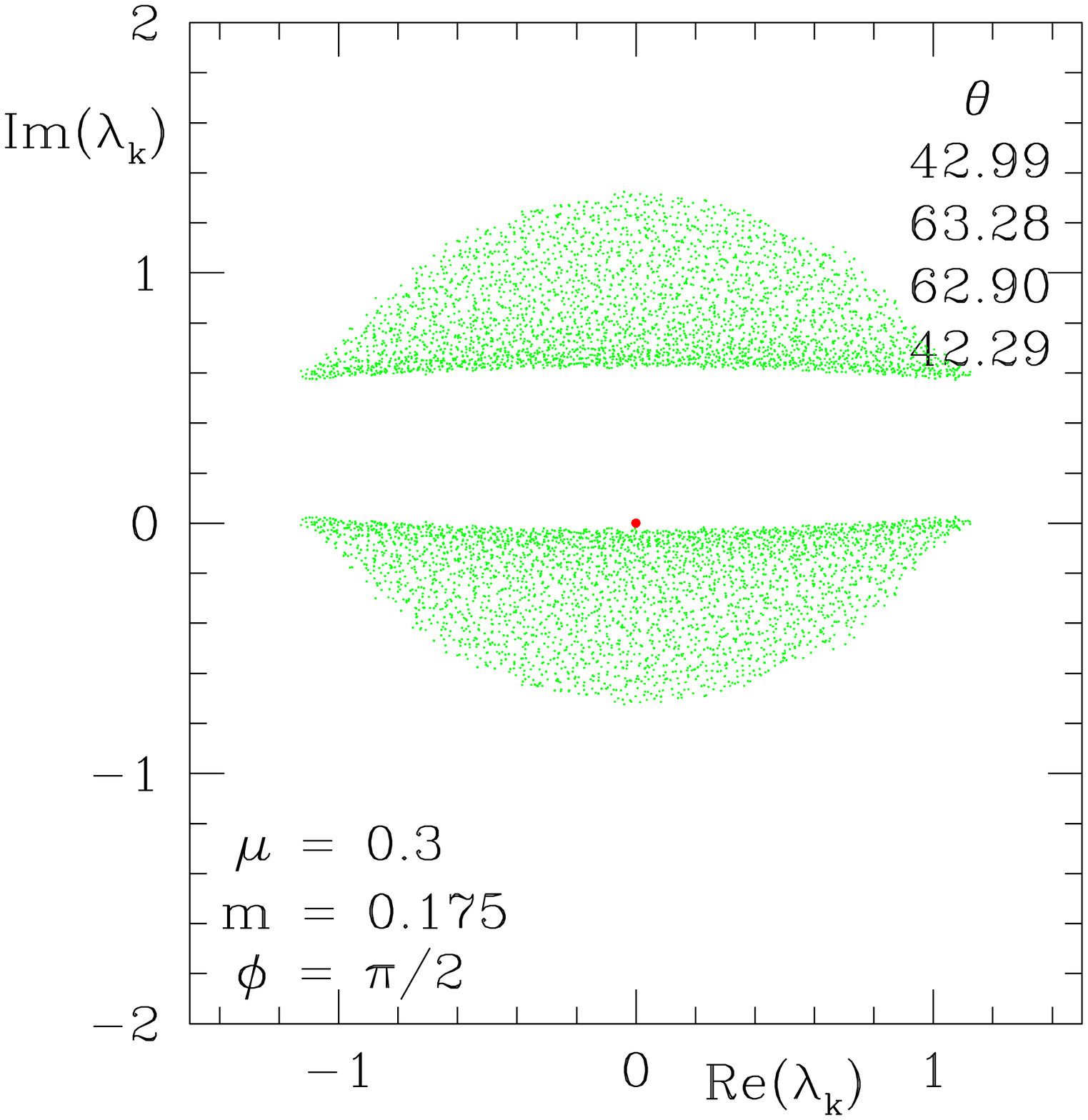,clip=,width=7cm}
\caption{\label{fig:scat}Scatter plot of the spectrum of a 
Random Matrix Dirac operator for $\mu=m_\pi/2$; 
{\bf left:} the eigenvalues of $D+\mu\gamma_0+m$ 
{\bf right:} the eigenvalues of $i(\gamma_0(D+m)+\mu)$.
In both cases the support of the spectrum has reached the origin indicated by the red point. Beyond this point, ie.~for 
$\mu>m_\pi/2$ the distribution of the chiral condensate and the baryon
has power law tails. (A similar phenomenon is expected to happen for 
lattice QCD with Wilson fermions in the Aoki phase \cite{DSV}.)} 
\end{figure}

\section{The distribution of $n$ from Chiral Perturbation Theory}

Despite the fact that pions have zero baryon charge the distribution
of the baryon number over the gauge fields is non trivial when computed
within Chiral Perturbation Theory.  
Certainly in Chiral Perturbation Theory we have that 
\begin{equation}
\frac{1}{2}\frac{1}{Z_{1+1}} \ \frac{d}{d\mu}  \ Z_{1+1} = \langle n \rangle = 0 \quad {\rm and} \quad 
\frac{1}{Z_{1+1}} \ \frac{d^2}{d\mu^2}  \ Z_{1+1}  = 0,
\end{equation}
but the average of the square of $n$ is non zero 
\be 
\langle n^2 \rangle = 
 \left.\frac {d^2} {d\mu_1d\mu_2} G_0(\mu_1, \mu_2)
 \right |_{\mu_1 = \mu_2 =\mu} \neq 0, 
\ee
since the 1-loop free energy is 
\begin{equation}
G_0(\mu_1,\mu_2) = V\frac{m_\pi^2T^2}{\pi^2}\sum_{n=1}^\infty \frac{K_2(\frac{m_\pi n}{T})}{n^2}
\cosh(\frac{\mu_1-\mu_2}{T}n). \nn
\end{equation}
So $\langle \delta(n-n') \rangle$ must necessarily be non trivial in
Chiral Perturbation Theory.

In order to compute the full distribution $P_{n}(x,y)$ it is necessary
to evaluate all moments 
\be
\langle\Ree[n]^k\Imm[n]^j\rangle
\ee
 in Chiral
Perturbation Theory. The details are given in \cite{LSV2} and involve  
an interesting combinatorial use of the replica trick
\cite{replica}. For the computation it is essential to specify whether
the chemical potential is larger or smaller than $m_\pi/2$, since in
the replicated generating functions additional condensates appear at
this scale. Also from the perspective of the eigenvalues of the Dirac
operator it is clear that the case $\mu<m_\pi/2$ must be very 
different from the one with $\mu>m_\pi/2$, see figure \ref{fig:scat}.
\begin{center}
\begin{figure}
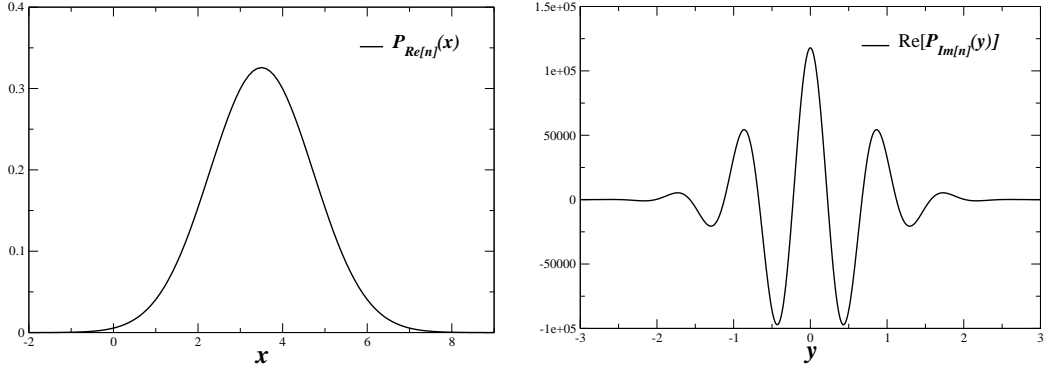

\centerline{\includegraphics[width=6.5cm,angle=0]{Pre.eps} \hspace{3mm}
\includegraphics[width=6.8cm,angle=0]{Pim.eps}}
\caption{\label{fig:PrePim} The distribution of the baryon number over
the gauge fields in the grand canonical ensemble. For $\mu<m_\pi/2$ the
distribution factorizes into the distribution of the real part (left
figure) and the imaginary part (right figure). The
distribution of the imaginary part takes complex values, shown is the
real part. Note the difference in the scales on the vertical axis. The
amplitude of the distribution of the imaginary part of $n$ grows
exponentially with the volume.} 
\end{figure}
\end{center}
\vspace{-5mm}

We first discuss the distribution of $n$ for $\mu<m_\pi/2$. To one-loop
order in Chiral Perturbation Theory the distribution factorizes \cite{LSV2}
\be
\label{PnB}
P_{n}(x,y) = 
P_{\Ree[n]}(x)P_{\Imm[n]}(y),
\ee
where the two factors take simple Gaussian forms
\be
P_{\Ree[n]}(x) & = & \frac{1}{\sqrt{\pi(\chi_{ud}^B+\chi_{ud}^I)}}
e^{-{(x-\nu_I)^2}/{(\chi_{ud}^B+\chi_{ud}^I)}} \\
P_{\Imm[n]}(y) & = & \frac{1}{\sqrt{\pi(\chi_{ud}^I-\chi_{ud}^B)}}
e^{{-(y-i\nu_I)^2}/{(\chi_{ud}^I-\chi_{ud}^B)}}. 
\ee
Note in particular that $P_{\Imm[n]}(y)$ takes complex values. It is
quite natural that the sign problem manifest it self in the
distribution of the imaginary part of $n$ since 
\be
\label{1alt}
n & \equiv &  
 \frac{d}{d\mu} \log\det(D+\mu\gamma_0+m) 
=  \frac{d}{d\mu} \log
 |{\det}(D+\mu\gamma_0+m)| + i\frac{d}{d\mu} \theta . 
\ee
In the above expression for the distribution of $n$ we have made use
of the notation
\be 
\nu_I &\equiv& \left . 
\frac d {d\mu_1} \Delta G_0(\mu_1, -\mu) 
\right|_{\mu_1 = \mu} \\
\chi^B_{ud} & \equiv&  \left .
 \frac {d^2} {d\mu_1d\mu_2} \Delta G_0(\mu_1, \mu_2)
 \right |_{\mu_1 = \mu_2 =\mu} \nn \\
\chi^I_{ud} & \equiv&  \left . 
 \frac {d^2} {d\mu_1d\mu_2} \Delta G_0(-\mu_1, \mu_2)
 \right |_{\mu_1 = \mu_2 =\mu}, \nn
\ee
where the free energy difference is (note that
 $\chi_{ud}^I+\chi_{ud}^B>0$ and $\chi_{ud}^I-\chi_{ud}^B>0$)
\begin{equation}
\Delta G_0(\mu_1,\mu_2) = V\frac{m_\pi^2T^2}{\pi^2}\sum_{n=1}^\infty \frac{K_2(\frac{m_\pi n}{T})}{n^2}
\left[\cosh(\frac{\mu_1-\mu_2}{T}n) - 1\right].
\end{equation}
Since all of the above quantities are extensive the amplitude
of $P_{\Imm[n]}(y)$ grows exponentially with the volume, the width
grows like $\sqrt{V}$, while the period of the oscillations are of
order $V^0$. For a plot of the distribution see figure
\ref{fig:PrePim}. 
The extreme oscillations of $P_{\Imm[n]}(y)$ are essential in order to
obtain zero expectation value of the quark number in Chiral
Perturbation Theory
\be
\langle n\rangle & = & \int dx dy \ (x+iy) P_{n}(x,y)  \\
 & = &  \int dx \ x P_{\Ree[n]}(x) + i \int dy \ y P_{\Imm[n]}(y) \nn \\
 & = & \nu_I + i i\nu_I = 0 \nn 
\ee 
The detailed cancellation between the contribution from the real part and
the imaginary part is only possible if the phase of the fermion the
determinant is accounted for properly. Similar
cancellations also take part for the higher moments of the baryon
number as well as for the moments of the chiral condensate, see
\cite{LSV2} for details.

\section{Complex Langevin}

One use of the results for the distribution of the baryon number is 
to illustrate how the complex Langevin method can deal with sign
problems in simple models. Clearly the distribution of the imaginary
part of the baryon number over the gauge fields is the challenging
part. Therefore, let us ask if complex Langevin is able to do the
simple one dimensional integral
\be
\int dy \ y \ P_{\Imm[n]}(y),
\ee
that is, to measure the contribution to the average baryon number from
the imaginary part of $n$.

To this end we define the complex Langevin action for $y=\Imm[n]$ as
\be
 S = -\log[P_{\Imm[n]}(y)] 
=-(iy+\nu_I)^2/(\chi_{ud}^I-\chi_{ud}^B) .
\ee
The next step is to complexify $\Imm[n]$ as $y=a+ib$ and write
down the flow equations for $a$ and $b$
\be
 a_{n+1} & = &  a_n - \epsilon \frac{2a_n}{\chi_{ud}^I-\chi_{ud}^B} +
\sqrt{\epsilon} \eta_n \\
  b_{n+1} & = &  b_n - \epsilon
  \frac{2(b_n-\nu_I)}{\chi_{ud}^I-\chi_{ud}^B}. \nn 
\ee
Note that the flow equations decouple. The equation for $a$ is that of
a Gaussian for which complex Langevin works perfectly. That of $b$ 
simply shifts $y$ by $\nu_I$ in the imaginary direction. Since there 
is no noise in the imaginary direction, the complex Langevin method 
effectively 
shifts the contour of the $y$-integral by a term of order $V$ in the
imaginary direction. After the shift, a simple integral over a Gaussian
without oscillations is left and the complex Langevin method has no problem in
evaluating this. Clearly the shift of the contour is the only reasonable
thing to do in this case, the strength of the complex Langevin
method is that it can make this shift automatically. A similar example was
worked out in \cite{Aarts} and \cite{PdF}.

For $\mu>m_\pi/2$ the chemical potential enters the spectral support of 
$\gamma_0(D+m)$ and the distribution of the baryon number develops
power law tails \cite{LSV2}. Nevertheless, complex Langevin is also 
able to deal with the sign problem for one dimensional QCD \cite{AS}
in this region. 

\section{Summary}

The interplay between lattice QCD and analytical studies of QCD is 
essential to understand 
QCD at nonzero chemical potential. Due to the sign problem, the
standard methods of lattice QCD only have a limited range of
applicability. In order to study dense strongly interacting matter from
first principles new numerical methods must be invented and put to
use. To understand how such methods can be designed it is essential to
understand how the sign problem affects physical observables such as
the baryon number and the chiral condensate.

Here we have derived the distribution of the baryon number over the
gauge fields from Chiral Perturbation Theory. We have shown that the
distribution takes complex values and is strongly oscillating. These
oscillations were shown to be central to the detailed cancellations 
which take place when forming the average baryon number. The
distributions also give detailed information on the overlap problem as
will be discussed in \cite{LSV3}. Here we have used the distribution
of the baryon number to show how the complex Langevin method can deal 
with sign problems. An important point to take away from this is
that the complex Langevin method works equally well independent of the 
volume and hence independently of the strength of the sign
problem. Similarly, in the  
well known cases \cite{Amb} where the complex Langevin method fails it
does so independently of the volume.

It is also possible to compute the distribution of the baryon number
{\sl over the phase of the fermion determinant} within Chiral Perturbation
theory \cite{LSV1}. Also in this case the complex and
oscillating nature of the distribution is essential in order to obtain
the correct physics at nonzero chemical potential. That
calculation also directly demonstrates that all phases of the fermion
determinant are important.

\end{document}